\documentstyle[12pt]{article}
\input{psfig.sty}
\newcommand{\pom}{I\!\! P}
\def\met{\not\!\!E_T}
\input{psfig.sty}
\begin{document}
\vspace*{-6ex}
{\hfill

\begin{center}

{\large\bf{Diffraction Results from CDF\footnote{Presented at {\em XIX International Workshop on Deep-Inelastic Scattering and Related Subjects (DIS 2011)}, April 11-15, 2011, Newport News Marriott at City Center, Newport News, VA USA.}}}

%\vglue 0.6cm
\vglue 0.3cm
{KONSTANTIN GOULIANOS\\ (for the CDF~II Collaboration)}
\vglue 0.2 cm
\baselineskip=13pt
{\em The Rockefeller University, 1230 York Avenue, New York, NY 10065, USA\\}
\vskip0.25in

%\today
%\vglue 0.5cm

\end{center}

\centerline{ABSTRACT}
We present final results by the CDF~II collaboration on diffractive $W$ and $Z$ production, report on the status of ongoing analyses on diffractive {\em dijet} production and on rapidity gaps between jets, and briefly summarize results obtained on exclusive production pointing to their relevance to calibrating theoretical models used to predict exclusive Higgs-boson production at the LHC.  
%{\rightskip=10pc
%\leftskip=10pc

}
\section{Introduction}
Starting with the first $\bar pp$ collider data at the Tevatron in 1989, the CDF Collaboration has been carrying on a comprehensive diffractive physics program aimed at understanding the QCD mechanism of diffraction. 
It is presumed that in diffraction a strongly-interacting color-singlet quark/gluon combination with the quantum numbers of the vacuum (the {\em Pomeron}, $\pom$) is exchanged~\cite{Collins}-\cite{Donnachie}. The aim of diffractive studies is to decipher the parton distribution function (PDF) of $\pom$ exchange. There is also a practical reason for diffractive studies. As approximately one quarter of all inelastic $\bar pp$ collisions at Tevatron energies are diffractive, they have a significant effect on the underlying event (UE) of hard (high transverse momentum) processes. Therefore, understanding diffraction can provide a tool for all data analyses where the UE influences trigger efficiencies and acceptance corrections. Since no radiation is expected from vacuum exchange, a large non-exponentially-suppressed pseudorapidity region devoid of particles, called a {\em rapidity gap}~\cite{rapidity}, is produced and can serve as an experimental signature for diffraction. Depending on the dissociation pattern, diffractive processes are classified as single-dissociation or single-diffraction (SD~-with one forward gap adjacent to a surviving $p$ or $\bar p$), double-dissociation or double-diffraction (DD-~with one central gap), and central-dissociation or double-Pomeron exchange (CD or DPE~-with two forward gaps). 

In Run~I, CDF studied all soft/inclusive (SD, DD, CD) and several hard ($W$, $dijet$, $J/\psi$, and $b$-quark) diffraction processes using the rapidity-gap signature to select diffractive events, and in some cases a Roman Pot Spectrometer (RPS) to measure the momentum of the surviving $\bar p$. While all Run~I results were found to be self-consistent within the RENORM model~\cite{renorm}, based on a renormalized Regge phenomenology to account for overlapping rapidity gaps, there were two striking disagreements with other experiments. First, depending on the model used for estimating gap acceptance/survival, the D0 Collaboration measured a larger fraction of SD to ND $W$ events by a factor of up to $\sim 3.5$; and second, CDF measured a ratio of diffractive to non-diffractive (ND) structure functions that was $\sim 20$\% greater than expectations based on HERA $ep$ measurements. To address these issues, special forward detectors were built and commissioned in Run~II. The forward detectors were also used to make a series of measurements on exclusive production of specific final states relevant to diffractive Higgs-boson production at the large Hadron Collider (LHC).

The status of the CDF analyses and final/preliminary results on diffractive and exclusive production are presented in Sec.~\ref{sec:results}, and conclusions are drawn in Sec.~\ref{sec:conclusion}.
\section{Results\label{sec:results}}
In this section we present final results for $W/Z$ production (Sec.~\ref{sec:WZ}) and preliminary results of the {\em dijet} (Sec.~\ref{sec:dijet}) and {\em gaps between jets} (Sec.~\ref{sec:gaps}) analyses.
\vspace*{-0.25cm}

\subsection{Diffractive $W$ and $Z$ production}\label{sec:WZ}
This analysis was fully reviewed in {\em DIFFRACTION 2010}~\cite{WZd2010}. Here, we present final results~\cite{WZprd} for events in the regions of $\bar p$ momentum-loss fraction,  $\xi$,   within $0.01<\xi<0.10$, and 4-momentum-transfer squared,  $t$,   within $-1<t<0$ (GeV/$c$)$^2$. 

Figure~\ref{fig1} shows LO diffractive $W$ and $Z$ production diagrams. The results are:
\vspace{-0.25cm}
\begin{figure}[h]\label{fig1}
%\hspace*{0.25cm}\includegraphics[height=.145\textheight]{fig1.eps}
\hspace*{0.25 cm}\psfig{figure=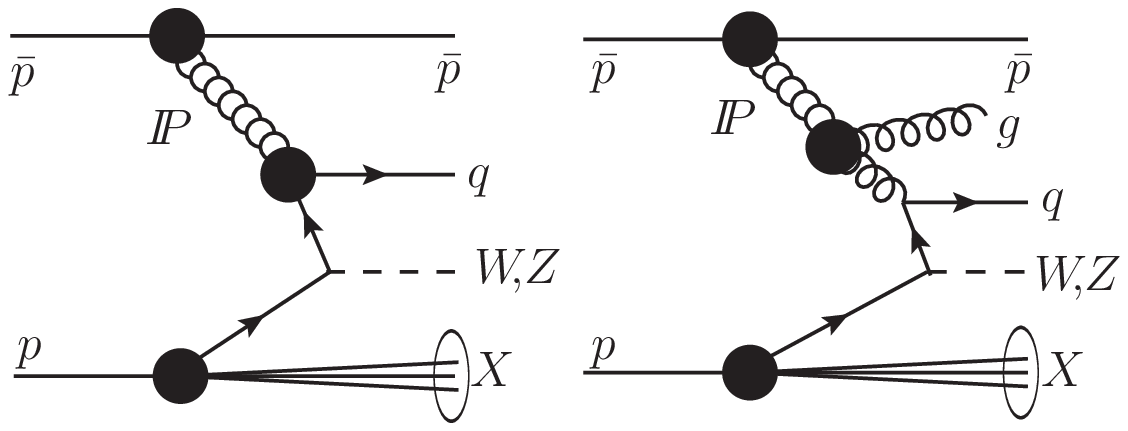,height=.125\textheight}
%\hspace*{-0.5cm}\includegraphics[height=.25\textheight]{fig2.eps}
\hspace*{-0.5cm}\psfig{figure=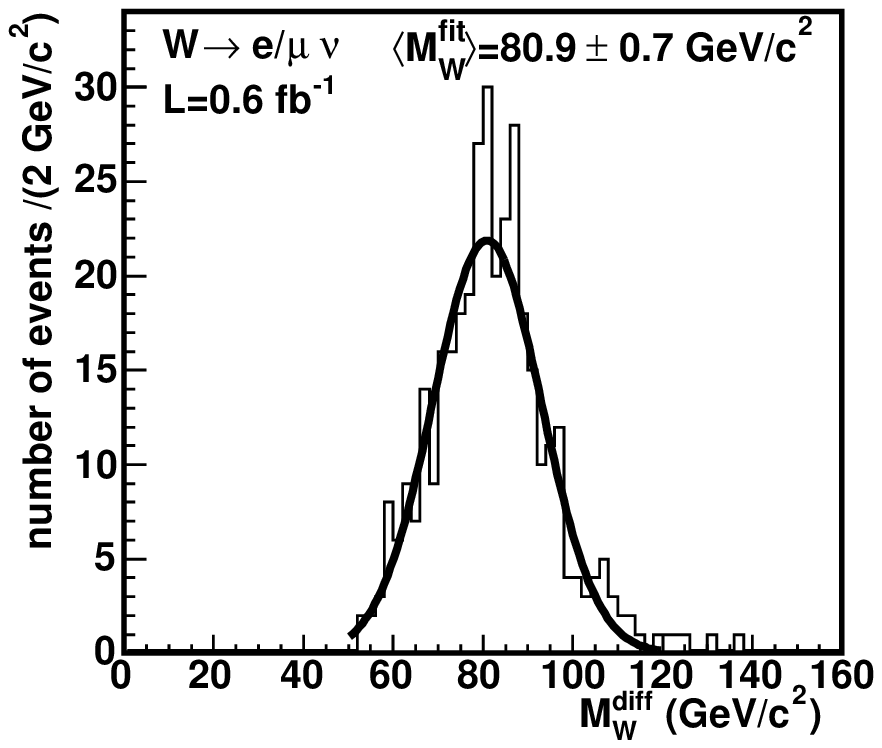,height=.24\textheight}
\caption{Diffractive $W$ and $Z$ production diagrams and ${\rm M}_W^{\rm diff}$ from diffractive $W$ events.}
\end{figure}
\begin{itemize}
\item SD/ND ratios for SD events within $0.03<\xi<0.10$ and   $ -t <0$ (GeV/$c$)$^2$:
\protect{$${\rm R}^{\rm sd/nd}_{\rm W}=[1.00 \pm 0.05\,(\mbox{stat.}) \pm 0.10(\,\mbox{syst.})], {\rm R}^{\rm sd/nd}_{\rm Z}=[0.88 \pm 0.21(\,\mbox{stat.}) \pm 0.08(\,\mbox{syst.})]\%.$$}
\noindent The ${\rm R}^{\rm sd/nd}_{\rm W}$ value confirms the CDF Run~I rapidity-gap-based result~\cite{run1W}. 
\item ${\rm M}_{\rm W}$ is measured from fully reconstructed diffractive $W$ events by obtaining $p_z^\nu$ for $W\rightarrow \mu/e+\nu$ from the difference between $\xi_{\bar p}^{\rm RPS}$ and its calorimetric value $\xi_{\bar p}^{\rm CAL}$:
\protect{$$\label{forward_mom}\xi^{\rm CAL}_{\bar p}=\sum_{i=1}^{N_{\rm towers}}\frac{E^i_{\rm T}}{\sqrt{s}}e^{-\eta^i},\;
\xi^{\rm RPS}_{\bar p}-\xi_{\bar p}^{\rm CAL}=\sum_{i=1}^{N_{\rm towers}}\frac{\met^i}{\sqrt{s}}e^{-\eta^\nu},\;p_z^\nu=\met/\tan{\left[2\tan^{-1}{(e^{-\eta^\nu})}\right]}.$$}
\end{itemize}
The measured value of  ${\rm M}_{\rm W}^{\rm diff}=89.9\pm0.7$~GeV/$c^2$, shown in Fig.~\ref{fig1}~(right), agrees with the world average of ${\rm M}_{\rm W}^{\rm PDG}=(80.399\pm0.023)$~GeV/$c^2$~\cite{PDG}. 
\subsection{Diffractive {\em dijet} production\label{sec:dijet}}
 We discuss the status of two analyses: `` Measurement of the structure function in single-diffraction dijet production'' and ``Gaps between jets''.  
\subsubsection{Structure function is single-diffraction {\em dijet} production}
Substantial progress has been made in this analysis since {\em EDS2009}~\cite{eds2009Factorization}, but updated results have not yet been released. 
The main conclusions remain the same:
\begin{itemize}
\item the measured $x_{Bj}$ rates confirm the factorization breakdown observed in Run~I;
\item In the range  $10^2$~(GeV$/c$)$^2<Q^2<10^4$~(GeV$/c$)$^2$, where the inclusive $E_{\rm T}$ measured distribution falls by a factor of $\sim 10^4$, the ratio of the SD/ND distributions increases by only a factor of $\sim 2$.  
\item The slope parameter $b(Q^2)|_{t=0}$ of an exponential fit to $t$ distributions near $t=0$ shows no $Q^2$ dependence in the range 1~(GeV$/c$)$^2<Q^2<10^4$~(GeV$/c$)$^2$.
\end{itemize}
These results support a picture of a composite Pomeron formed from color-singlet combinations of the underlying parton densities of the nucleon (see, e.g.,~\cite{renorm}). 

Currently, we are working on  extending the measurement of the $t$ distribution to $t\sim -4$~(GeV$/c$)$^2$ to search for a diffraction minimum. 
\subsubsection{Gaps between jets\label{sec:gaps}}
An update of this analysis has been recently presented in~\cite{WZd2010}. Jet-Gap-Jet (JGJ) event rates can be used to test perturbative gap-creation models, such as the BFKL hypothesis~(see, e.g.,~\cite{Donnachie}). To reduce model dependence, we measure ratios of gap events to all events, ${\rm R}_{\rm gap}\equiv {\rm N}_{\rm gap}/{\rm N}_{\rm all}$, as a function of the width of the gap and study the suppression relative to expectations between JGJ and soft DD events selected by their activity in the MiniPlugs in the $\eta$-range $3.5<|\eta|<5.1$. 
 We find that the ${\rm R}_{\rm gap}^{\rm jet}$ ratios are suppressed relative to ${\rm R}_{\rm gap}^{\rm DD}$, as expected, but the suppression is independent of the width of the gap. A BFKL-model contribution to the JGJ distribution would be expected to be concentrated at high $\Delta\eta$. No excess that could be attributed to a BFKL contribution is observed.
\subsection{Exclusive production\label{sec:exclusive}}
The main interest in studying diffractively produced exclusive final states is to use the results to check/calibrate QCD models of diffraction that can be applied to calculate production rates of exclusive $Higgs$ production at the LHC. Final states studied include $JJ$ ({\em dijet})~\cite{exclJJ2008}, $\chi_c$~\cite{exclchic2007}, $\gamma\gamma$~\cite{exclgamma2007}, and $J/\psi$ and $\psi(2s)$~\cite{exclJpsi2009}. The results are in good agreement with the model of~\cite{KMRexclusive}.

\section{Conclusion\label{sec:conclusion}}
 We present final results by the CDF~II collaboration on diffractive $W$ and $Z$ production and report on the status of ongoing analyses on diffractive {\em dijet} production and on rapidity gaps between jets.  

The diffractive $W/Z$ analysis has been completed and the results are published~\cite{WZprd}. We find that in the range of $\bar p$ forward momentum loss $0.03<\xi_{\bar p}<0.10$ and for $-1<t<0$ (GeV/$c$)$^2$ the fraction of diffractive events in $W$ and $Z$ production is ${\rm R}_{\rm W}=[1.00 \pm 0.05\,(\mbox{stat.}) \pm 0.10(\,\mbox{syst.})]\%$ and  ${\rm R}_{\rm Z}=[0.88 \pm 0.21(\,\mbox{stat.}) \pm 0.08(\,\mbox{syst.})]\%$, respectively. The $R_W$ value is compatible with our Run~I rapidity-gap based result. 

In the analysis on the diffractive structure function in {\em dijet} production, we are working to extend the measurement of the $t$ distribution to $t\sim -4$~(GeV$/c$)$^2$ to search for a diffraction minimum; and in the {\em gaps between jets} analysis, we are reanalyzing the data to obtain results in a format more suitable for comparison with theoretical predictions.

We also summarize results on exclusive production, pointing to their relevance to calibrating theoretical models used to predict exclusive {\em Higgs} production at the LHC. 
\section*{Acknowledgments}
I would like to thank my colleagues at The Rockefeller University and all members of the CDF Collaboration for their invaluable contributions to this project, and the Office of Science of the Department of Energy for providing financial support.  
%\end{theacknowledgments}


\begin{thebibliography}{9}
\bibitem{rapidity} Rapidity, $y=\frac{1}{2}\ln\frac{E+p_L}{E-p_L}$, and pseudorapidity, $\eta=-\ln\tan\frac{\theta}{2}$, where $\theta$ is the polar angle of a particle with respect to the proton beam ($+\hat z$ direction), are used interchangeably for particles detected in the calorimeters, 
since in the kinematic range of interest in this analysis they are approximately equal.
\end{thebibliography}
\end{document}